\documentclass[twocolumn,showpacs,amsmath,amssymb,superscriptaddress,prl]{revtex4}

\usepackage{graphicx}
\usepackage{dcolumn}
\usepackage{bm}

\begin{document}

\title{Full Counting Statistics in Strongly Interacting 
	Systems: Non-Markovian Effects}

\author{Alessandro Braggio}

\affiliation{Institut f\"ur Theoretische Physik III, Ruhr-Universit\"at Bochum,
44780 Bochum, Germany}

\affiliation{Dipartimento di Fisica, Universit\`a di Genova, INFM-LAMIA, 
Via Dodecaneso 33, 16146 Genova, Italy}

\author{J\"urgen K\"onig}

\affiliation{Institut f\"ur Theoretische Physik III, Ruhr-Universit\"at Bochum,
44780 Bochum, Germany}

\author{Rosario Fazio}

\affiliation{NEST-INFM \& Scuola Normale Superiore, Piazza dei Cavalieri 7, 
56126 Pisa, Italy}

\date{\today}

\begin{abstract}

We present a theory of full counting statistics for electron transport through
interacting electron systems with non-Markovian dynamics.
We illustrate our approach for transport through a single-level 
quantum dot and a metallic single-electron transistor to second order in 
the tunnel-coupling strength, and discuss under which circumstances 
non-Markovian effects appear in the transport properties.
\end{abstract}


\maketitle


The study of current fluctuations in mesoscopic systems has become an intense 
field of research, since it allows to access information about electron 
correlations that is not contained in the average current.
The phenomenon of shot noise~\cite{Kogan96,BlanterButtiker99,NazarovQN03}, 
that dominates the current noise at low temperatures, has been investigated 
theoretically and experimentally in various contexts. 
Further understanding in electron transport can be gained from the study of 
higher moments~\cite{Reulet03} which can be conveniently extracted from the 
study of full counting statistics (FCS) \cite{Levitov93-96,Nazarov99}. 
Up to date FCS has been studied in a variety of cases. Examples include 
normal-superconductor hybrid structures~\cite{Muzykantskii94}, 
superconducting weak links~\cite{Belzig01SS}, tunnel 
junctions~\cite{ShelankovRammer03}, chaotic cavities~\cite{Jong96}, 
entangled electrons~\cite{Taddei02-04} and spin-correlated 
systems~\cite{DiLorenzo03}. 
Schemes for an experimental measurement of FCS has been put forward in 
Refs.~\onlinecite{FCSmeasure}.

Electron-electron interaction may strongly influence quantum transport. 
The effect of electron correlation on FCS has been considered so far in 
the case of a weakly interacting mesoscopic conductor~\cite{Kindermann03prl}, 
almost open dots~\cite{Andreev01,BagrestNazarov03O,KindermannTrauzettel04},
charge pumping~\cite{Andreev01} and charge shuttles~\cite{Pistolesi04}.
The FCS for Coulomb-blockade devices has been 
analyzed~\cite{BagretsNazarov03,Choi01,Makhlin00} in the framework of 
a Born-Markov master equation approach.
The aim of the present paper is to extend this idea to a obtain a
general theory of full counting statistics for strongly interacting systems 
with non-Markovian behavior.
In particular, we formulate a perturbative non-Markovian expansion that allows
for a systematic study of the relative importance of non-Markovian corrections.
We demonstrate for the example of a single-level quantum dot with strong
Coulomb interaction that non-Markovian effects become increasingly important
for higher moments of the current fluctuations.

{\em Full counting statistics.} --
Full information about all transport properties of a given system is contained 
in the probability distribution $P(N,t)$ that $N$ charges have passed through 
the system after time $t$.
This distribution function is related to the cumulant generating function 
$S(\chi)$ by
\begin{equation}
\label{cgf}
  S (\chi) = - \ln \left[ \sum_{N=-\infty}^\infty e^{iN\chi}P(N,t) \right] 
  \, , 
\end{equation}
where $\chi$ is the counting field.
All moments of the current can be obtained from the cumulant generating 
function by performing derivatives with respect to the counting field 
$\langle \langle I\rangle \rangle_n=-(-i)^n( e^n/t) \partial^n_\chi S(\chi)|_{\chi=0}$.
The first four moments are the average current, the (zero-frequency) current 
noise, the skewness, and the kurtosis.

In this work we consider systems with strong local interactions, such as electrons in a 
quantum dot, that are coupled to a reservoir of noninteracting degrees 
of freedom. In these situations transport properties can often be described in terms of 
few (local) degrees of freedom (the charge of the quantum dot in the previous 
example). It is then convenient to integrate out the noninteracting degrees of 
freedom to arrive at an effective description of the reduced system only.
Let $\mathbf{p}^{\rm in}$ be the vector of probabilities to find the system in 
the corresponding state at the initial time $t=0$.
The time evolution of the system is described by a generalized master equation,
\begin{equation}
\label{master}
  \frac{d}{dt}\mathbf{p}(N,t) = \sum\limits_{N'=-\infty}^\infty
  \int\limits^{t}_{0} d t' \,\, \mathbf{W}(N-N',t,t')\cdot \mathbf{p}(N',t') 
  \, ,
\end{equation}
where $\mathbf{p}(N,t)$ is the vector of dot occupation probabilities
under the condition that $N$ electrons have passed the system.  The
cumulant generating function is given by Eq.~(\ref{cgf}) where $P(N,t)
= \mathbf{e}^T \cdot \mathbf{p}(N,t)$ with $\mathbf{e}^T = (1,1,\ldots
,1)$.  The matrix $\mathbf{W}(N-N',t,t')$ describes transitions during
which $N-N'$ electrons are transferred.  The counting field $\chi$ is
introduced by Fourier transforming the master equation
$\mathbf{p}(\chi,t) = \sum_N \exp(iN\chi)
\mathbf{p}(N,t)$ and $\mathbf{W}(\chi,t,t') = \sum_N \exp(iN\chi) 
\mathbf{W}(N,t,t')$.

In general, the kernels $\mathbf{W}(\chi,t,t')$ are nonlocal in time as a
consequence of having integrated out the reservoir degrees of freedom. 
We consider the case in which there is no explicit time dependence of the 
systems parameters, so that $\mathbf{W}(\chi,t-t')$ can be Laplace transformed,
$\mathbf{W}(\chi,z) = \int_0^{\infty} d t \, \exp(-zt) \mathbf{W}(\chi,t)$.
Usually, the dynamics of the system is characterized by a time scale of the 
individual transitions. We assume that $\mathbf{W}(\chi,t)$ decays faster 
than any power of $t$, such that the derivatives 
$\partial_z^n \mathbf{W}(\chi,z)$ exist for all $n$. 

In the Markovian limit, $\mathbf{W}(\chi,t) \sim \delta (t)$, the system at 
time $t$ depends only on the state of the system at the same time $t$. 
The master equation Eq.~(\ref{master}) is,
then, solved by $\mathbf{p}(\chi,t) = \exp[\mathbf{W}(\chi,z=0) t]
\cdot \mathbf{p}^{\rm in}$, i.e., only the $z=0$ component of the
transitions $\mathbf{W}(\chi,z)$ are taken into account.

Our goal is to describe stationary transport properties in the presence of a 
memory of the system, described by the full $z$-dependence of
$\mathbf{W}(\chi,z)$. 
To solve the master equation without making use of the Markovian approximation,
we switch to the Laplace representation,
\begin{equation}
  z \mathbf{p}(\chi,z) - \mathbf{p}^{\rm in} = 
  \mathbf{W}(\chi,z) \cdot \mathbf{p}(\chi,z) \, ,
\end{equation}
which is solved by $\mathbf{p}(\chi,z) = \sum_{n=0}^\infty 
[\mathbf{W}(\chi,z)]^n / z^{n+1} \mathbf{p}^{\rm in}$. 
By assuming that the kernel $\mathbf{W}(\chi,t)$ decays in time faster 
than any power law, we can define the Taylor series 
$[\mathbf{W}(\chi,z)]^n = \sum_{m=0}^\infty
\partial_z^m [\mathbf{W}(\chi,z)]^n \big|_{z=0} \, z^m /m!$ and substitute
it in the previous solution for $\mathbf{p}(\chi,z)$.
The long-time behavior of $\mathbf{p}(\chi,z)$ is determined by its poles in 
$z$.
Performing the inverse Laplace transformation we get
\begin{equation}
  \label{pchi}
  \mathbf{p}(\chi,t) =  \sum_{n=0}^\infty \frac{
  \partial_z^n \left( [\mathbf{W}(\chi,z)]^n {\rm e} ^{\mathbf{W}(\chi,z) t}
  \right)}{n!} \Bigr|_{z=0^+} \cdot \mathbf{p}^{\rm in} \, ,
\end{equation}
for large $t$.
To proceed, we perform a spectral decomposition of the matrix 
$\mathbf{W}(\chi,z)$.
For physical reasonable systems, all eigenvalues have a negative real part.
As a consequence of the exponential function in Eq.~(\ref{pchi}), the 
long-time behavior will be dominated by the eigenvalue $\lambda(\chi,z)$ 
with the smallest absolute value of the real part.
Let $\mathbf{q}_0$ and $\mathbf{p}_0$ be the corresponding left and right
eigenvectors, $\mathbf{q}_0^T \cdot \mathbf{W}(\chi,z) = \lambda(\chi,z) 
\mathbf{q}_0^T$, and $\mathbf{W}(\chi,z) \cdot \mathbf{p}_0 = \lambda(\chi,z) 
\mathbf{p}_0$. 
Unitarity in the absence of counting fields implies $\lambda(0,z)=0$ for all 
$z$.

The cumulant generating function becomes $ S (\chi) = -\ln \left[
\sum_{n=0}^\infty {1\over n!} \partial^n_z \left( \lambda^n {\rm
e}^{\lambda t + \mu } \right) \right]_{z=0^+} \, , $ with $\mu(\chi,z)
= \ln [(\mathbf{e}^T \cdot \mathbf{p}_0) (\mathbf{q}_0^T \cdot
\mathbf{p}_{\rm in})]$.  By performing the time derivative and making
use of the relation $\sum_{n=0}^\infty {1\over n!} \partial^n \left( a
b^{n+1} \right) /
\sum_{n=0}^\infty {1\over n!} \partial^n \left( a b^{n} \right)
=\sum_{n=0}^\infty {1\over n!} \partial^n \left( b^{n+1} \right) / 
\sum_{n=0}^\infty {1\over n!} \partial^n \left( b^{n} \right)$
that holds for arbitrary functions $a$ and $b$, we arrive at the central 
result of this Letter,
\begin{equation}
\label{final}
  S (\chi) =- 
  {\sum_{n=0}^\infty {1\over n!} \partial^n_z 
    \left[ \lambda^{n+1}(\chi,z) \right] \over 
    \sum_{n=0}^\infty {1\over n!} \partial^n_z 
    \left[ \lambda^{n}(\chi,z) \right] } \bigg|_{z=0^+} t \, .
\end{equation}
The cumulant generating function 
depends only on the eigenvalue $\lambda(\chi,z)$. 
This result can be used as a starting point for a non-Markovian expansion,
$S(\chi) = \sum_{n=0}^\infty S_n(\chi)$ where $S_n(\chi)$ contains $n$
derivatives with respect to $z$ applied to $n+1$ factors of $\lambda$.
While $S_0(\chi)$ describes the Markovian limit, $S_n(\chi)$ is the $n$-th 
non-Markovian correction. 

{\em Perturbative non-Markovian expansion.} -- For many systems there
is a small parameter which allows for a perturbative analysis of all
transport properties.  In the examples to be discussed below this will
be the tunnel-coupling strength between the leads and the interacting
region (quantum dot or metallic island).  Then, $\lambda(\chi,z) =
\sum_{i=1}^\infty
\lambda^{(i)}(\chi,z)$, where the superscript $(i)$ indicates the order in the
small parameter.  The lowest-order transport properties are derived
from the lowest-order cumulant generating function $S^{(1)}(\chi) = -t
\lambda^{(1)}(\chi,z)|_{z=0^+}$, as found in Ref.~\cite{BagretsNazarov03}.  
Non-Markovian corrections, signaled by derivatives 
$\partial_z^k\lambda(\chi,z)$, do not enter in this limit.  
The highest derivative that enters in the evaluation of the $n$-th 
moment in $m$-th order perturbation theory is 
$\partial^k_z \lambda (\chi,z)|_{z=0}$ with $k = \min\{n,m\} -1$.  
As a consequence, non-Markovian behavior is probed only in the
second or higher moment combined with second or higher order in
perturbation theory.
The second-order contribution for example reads 
$S^{(2)}(\chi) = -t [\lambda^{(2)} (\chi,z) + \lambda^{(1)}(\chi,z) 
\partial_z \lambda^{(1)}(\chi,z)]_{z=0^+}$. 
The appearance of these derivatives in the noise of second-order transport 
through quantum dots has been also found in Ref.~\onlinecite{Thielmann04}.
In the remaining part of this Letter, we illustrate our approach with two 
examples.
We calculate the cumulant generating function for second-order transport 
through a single-level quantum dot and through a metallic single-electron 
transistor in the presence of strong Coulomb interaction.

{\em Single level QD.}-- 
The single-level quantum dot is described by the Hamiltonian, 
$H=H_{\rm L} + H_{\rm R} + H_{\rm D} + H_{\rm T}$.
The electrons in the noninteracting left and right leads are represented by 
$H_{\rm L}$ and $H_{\rm R}$, respectively, 
$H_{\rm D}=\epsilon \sum_\sigma c^\dagger_\sigma c_\sigma  
+ U n_\uparrow n_\downarrow$ describes the dot with level energy $\epsilon$
and charging energy $U$ for double occupation.
Tunneling is modeled by
$H_{{\rm T},r}=\sum_{\sigma}t_r a^\dagger_{rk\sigma} c_\sigma 
+ \textit{h.c.}$ with $r={\rm R},{\rm L}$, where we assume
the tunnel matrix element $t_r$ to be independent of momentum $k$ and spin 
$\sigma$.
The tunnel-coupling strength is characterized by the intrinsic linewidth
$\Gamma=\Gamma_{\rm L}+\Gamma_{\rm R}$ with $\Gamma_r=2\pi\rho_r |t_r|^2$ where
$\rho_r$ is the density of states in the leads.
An asymmetry of the tunnel couplings is parametrised by
$\gamma =4\Gamma_{\rm L}\Gamma_{\rm R}/\Gamma^2$.

To derive the kernels $\mathbf{W}$ of the generalized master equation, we 
make use of a diagrammatic real-time technique \cite{konigRTT96} for the time 
evolution of the reduced density matrix formulated on a Keldysh contour.
We introduce counting fields 
$\chi_r$ for tunneling through barrier $r$ into lead $r$ by the replacement
$t_r \rightarrow t_r \exp( i \chi_r )$ for tunnel vertices on the upper and
$t_r \rightarrow t_r \exp( - i \chi_r )$ on the lower branch of the Keldysh
contour with $\chi_{\rm L} = -\chi_{\rm R} = \chi/2$.

We consider the limit $U\rightarrow \infty$, in which double
occupancy of the dot is prohibited \cite{rem}, and obtain
\begin{equation}
  \label{So1U}
    S^{(1)} (\chi) = \frac{t \Gamma \tilde{f}(\epsilon)}{2\hbar}
    \left[1 -\sqrt{1 +
	\frac{2 \gamma \sum_{k} f_{(k)}(\epsilon) ({\rm e}^{ik\chi }-1)}
	{[\tilde{f}(\epsilon)]^2}} \right]
\end{equation} 
with $f_{(+)}(\omega)=[1- f_{\rm L}(\omega)]f_{\rm R}(\omega)$,
$f_{(-)}(\omega)=f_{\rm L}(\omega)[1- f_{\rm R}(\omega)]$, and
$\tilde{f}(\omega)=\sum_r\Gamma_r[1+f_r(\omega)]/\Gamma$, where
$f_r(\omega)$ is the Fermi function for lead $r$.
This result was previously obtained 
in \cite{BagretsNazarov03}. 
The second-order contribution,
$S^{(2)} (\chi) =  S^{(2)}_{\rm cot}(\chi)+ S^{(2)}_{\rm ren}(\chi)$,
consists of two terms. The first one,
\begin{equation}
  \label{So20}
  S^{(2)}_{\rm cot}(\chi) =
  -\frac{t \gamma \Gamma^2}{4\pi\hbar}\sum_{k=\pm} ({\rm e}^{ik\chi}-1)\int\!\! d\omega f_{(k)}(\omega) R(\omega-\epsilon) ,
\end{equation}  
with $R(\omega)={\rm Re}[1/(\omega+i0^+)^2]$,
describes cotunneling processes \cite{cotunneling}, and is in agreement with 
previous work about noise in cotunneling regime~\cite{SukhorukovBukardLoss}.
The counting-field dependence corresponds to a bidirectional Poisson 
statistics of a single barrier where the transition rates are substituted by 
the cotunneling rates of the quantum dot.
However, $S^{(2)} (\chi)$ contains a second contribution,
\begin{equation}
  \label{So2ren}
  S^{(2)}_{\rm ren}(\chi) = \partial_\epsilon \left[ S^{(1)}(\chi)\  
    {\rm Re}[\sigma(\epsilon)] \right]
\end{equation}
with $\sigma (\epsilon) = -\sum_r(\Gamma_r/2\pi) \int d\omega f_r(\omega)/
(\omega-\epsilon+i0^+)$. 
In the previous
formulas an high energy cut-off $E_{\rm c}$, of the order of charging energy,
has to be introduced in order to cure spurious divergences related to the fact
that we restricted the charge states to 0,1~\cite{SET}. 
The contribution $S^{(2)}_{\rm ren}(\chi)$ is also of second order in the
tunnel-coupling strength but obeys the same 
statistics as the first-order (sequential-tunneling) result, Eq.~(\ref{So1U}).
This suggests that there are two different types of second-order contributions
to transport. 
In addition to the usual cotunneling processes, there are corrections to
sequential tunneling due to quantum-fluctuation induced renormalization of the
system parameters.
From the form of Eq.~(\ref{So2ren}) we deduce a renormalization of 
level position and coupling strength given by 
$\tilde \epsilon = \epsilon + {\rm Re}\, \sigma(\epsilon)$ 
and $\tilde \Gamma = \Gamma [ 1 + \partial_\epsilon {\rm Re}\, 
\sigma(\epsilon)]$.

\begin{figure}[h]
\includegraphics[width=\linewidth]{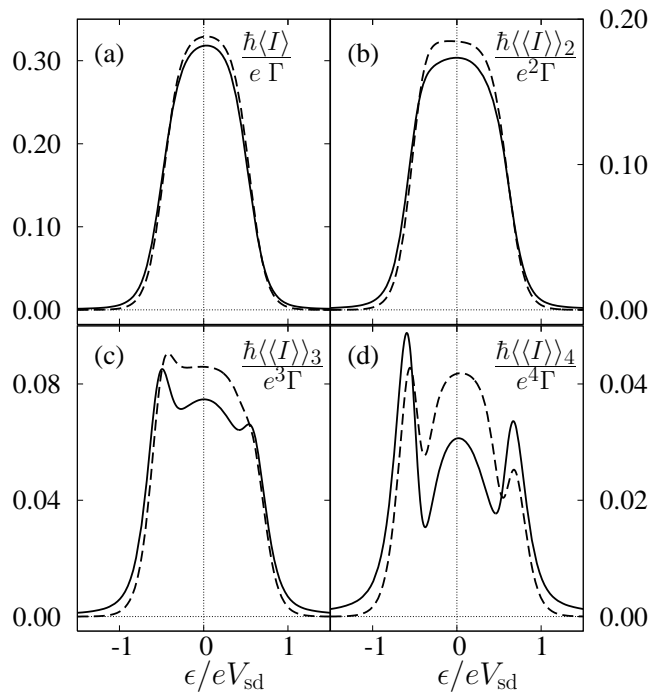}
  \caption{\label{fig1} 
    Current (a), noise (b), skewness (c), and kurtosis (d) for a single-level
    quantum dot with $U \rightarrow \infty$ as a function of the level position
    $\epsilon/eV_{\rm sd}$ for first (dashed lines) and first- plus 
    second-order (solid lines) in tunneling with 
    $eV_{\rm sd}=(\mu_{\rm R}-\mu_{\rm L})$. 
    Other parameters are $\gamma=1$, $\Gamma=3\cdot 10^{-2} eV_{\rm sd}$, 
    $k_{\rm B}T=10^{-1} eV_{\rm sd}$ and the high-energy cutoff 
    $E_{\rm c} = 10 eV_{\rm sd}$.
}
\end{figure}

In Fig.~\ref{fig1} we plot the first four moments (current, noise, skewness, 
and kurtosis) as a function of level position $\epsilon$. 
The solid lines represent the full first- plus second-order result, as compared
to the first-order contribution (dashed line).

\begin{figure}[h]
\includegraphics[width=\linewidth]{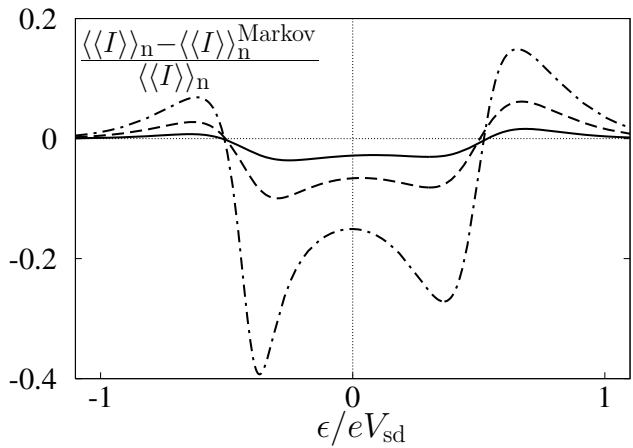}
  \caption{\label{fig2} 
    Relative contribution of the non-Markovian part 
    $\langle\langle I \rangle \rangle_{\rm n} -
    \langle \langle I\rangle \rangle_{\rm n}^{\rm Markov}$ to the
    $n$-th moment $\langle\langle I \rangle \rangle_{\rm n}$ in first- plus 
    second-order in tunneling.
    The full, dashed, and dot-dashed line corresponds to noise ($n=2$),
    skewness ($n=3$), and kurtosis ($n=4$) for the same parameters as in 
    Fig.~\ref{fig1}. 
  }
\end{figure}

The relative importance of the non-Markovian contributions is illustrated in 
Fig.~\ref{fig2}.
While for the current only Markovian contributions enter (see discussion 
above), non-Markovian corrections become increasingly important for higher
moments.

{\em Metallic QD.}-- 
A similar analysis can be performed for a metallic single-electron transistor,
which accommodates a continuum of states on the dot and includes a large number
of transverse channels.
Following the notation of Refs.~\onlinecite{SET}, we characterize the 
tunnel-coupling strength by the dimensionless conductance 
$\alpha_0^r=h/(4 \pi e^2 R_r)$ where $R_r$ is the resistance of barrier 
$r= \rm L,R$.
We concentrate on the low-temperature regime, in which only two charge states
of the metallic island have to be taken into account. 
This requires, again,
the introduction of an high-energy cut-off to regularize the integrals.
The difference of the charging energy between them is denoted by $\Delta$.
We obtain for the first order of cumulant generating function
\begin{equation}
  \label{So1UMe}
  S^{(1)}(\chi)=\frac{t\pi\alpha(\Delta)}{\hbar}
    \left[1-\sqrt{1+4 \sum_{k=\pm}
\frac{\alpha_{(k)}(\Delta)({\rm e}^{ik\chi }-1)}{\alpha(\Delta)} }\right]
\end{equation} 
where we have used the definitions 
$\alpha(\omega)=\sum_{r= \rm L,R}\alpha_r(\omega)$ and
$\alpha_{(\pm)}(\omega)=\alpha^\pm_{\rm L}(\omega)\alpha^\mp_{\rm R}(\omega)$
with $\alpha^\pm_{r}(\omega)=\alpha_r(\omega)f_r^\pm(\omega)$ where
$f_r^\pm(\omega)$ is the Fermi functions of lead $r$ and
$\alpha_r(\omega)=\alpha^r_0 \coth[\beta (\omega-\mu_r)/2]$.
The second-order term is given by $S^{(2)}(\chi)=S^{(2)}_{\rm cot}(\chi) + 
S^{(2)}_{\rm ren}(\chi)$, as the case of the single-level quantum dot.
Again, there is a contribution due to cotunneling,
\begin{equation}
  S^{(2)}_{\rm cot}(\chi) =- \frac{2\pi t}{\hbar} 
    \sum_{k=\pm} ({\rm e}^{ik\chi} -1)\int\!\! d\omega \alpha_{(k)}(\omega) R(\omega-\Delta) ,
\end{equation}
and a term
\begin{equation}
  S^{(2)}_{\rm ren}(\chi) = 
  \partial_\Delta\left[ S^{(1)} (\chi)\  {\rm Re}[\sigma(\Delta)]\right]
\end{equation}
associated with sequential-tunneling processes with renormalized system 
parameters, where $\sigma (\Delta) = -\sum_r\int d\omega \alpha_r(\omega)/
(\omega-\Delta+i0^+)$.
This interpretation of the different types of second-order contributions is
consistent with the analysis of the second-order current \cite{koenig_cot} and 
of the FCS within a drone Majorana fermion representation \cite{Utsumi05}. 

{\em Conclusions.} -- 
We present a theory of FCS for interacting systems with non-Markovian 
dynamics. 
A general expression for the cumulant generating function is derived 
that provides the starting point for a perturbative non-Markovian expansion.
As examples we study transport through a single-level quantum dot and a 
metallic single-electron transistor to second order in the tunnel-coupling 
strength.
From our formulation we could identify two different types of contributions
to second-order transport, namely cotunneling and corrections to sequential 
tunneling due to renormalization of the system parameters.
Furthermore, we demonstrate the increasing importance of non-Markovian effects
for higher moments and higher orders in the tunnel-coupling strength.

We thank D. Bagrets, W. Belzig, Y. Gefen, M. Hettler, G. Johansson, 
F. Plastina, A. Romito, M. Sassetti, and A. Thielmann, for useful 
discussions. Financial support from and 
DFG via SFB 491 and GRK 726, EU-RTN-RTNNANO, EU-RTN-Spintronics,
MIUR-Firb  is gratefully acknowledged.


\begin{thebibliography}{99}
\bibitem{rem}
We checked that for the opposite limit, $U=0$, our theory is consistent with 
the exact cumulant generating function obtained from Levitov's formula \cite{BagretsNazarov03}
and expanded up to second order on $\Gamma$.
\end{thebibliography}
\end{document}